\documentclass[twocolumn,showpacs,preprintnumbers,amsmath,amssymb]{revtex4}

\usepackage{graphicx}
\usepackage{dcolumn}
\usepackage{bm}

\def\bea{\begin{eqnarray}}
\def\eea{\end{eqnarray}}
\def\bi{\begin{itemize}}
\def\ei{\end{itemize}}

\begin{document}

\title{Stationary state in N-body System with power law interaction}%
\author{Osamu Iguchi}%
 \email{osamu@phys.ocha.ac.jp}%
 \affiliation{Department of Physics, Ochanomizu University, 
                2-1-1 Ohtuka, Bunkyo, Tokyo,112-8610 Japan}%

\begin{abstract}
Many self-gravitating systems often show scaling properties 
in their mass density, system size, velocities and so on.
In order to clarify the origin of these scaling properties, 
we consider the stationary state of N-body system 
with inverse power law interaction.
As a simple case, 
we consider the self-similar stationary solution 
in the collisionless Boltzmann equation with power law potential
and investigate its stability 
in terms of a linear symplectic perturbation.
The stable scaling solutions obtained are characterized 
by the power index of the potential 
and the virial ratio of the initial state.
It is suggested in general that 
the nonextensive system has much various stable scaling solutions 
than those in the extensive system.  
\end{abstract}

\pacs{05.20.Jj, 45.50.-j, 98.10.+z}

\maketitle


\section{\label{sec:INTRO}introduction}

There are many self-gravitating systems 
which are characterized by some scaling properties.
For example,
the inter stellar medium shows that 
its velocity dispersion $\sigma$ is power law related 
with the system size $L$ or mass $M$\cite{larson81} 
($\sigma\sim L^{0.38}\sim M^{0.2}$)
and isothermal contour is characterized 
by the fractal dimension $D\sim1.36$\cite{falgarone91}.
The observations by the Hubble Space Telescope show 
elliptical galaxies have a power law density distribution $\rho\sim r^{-n}$
(at outer region, $n\sim 4$ and 
at inner region, $n\sim 0.5-1.0$ for the bright elliptical galaxies and 
$n\sim 2$ for the faint ones\cite{merritt96}.). 
The distribution of the galaxies and the cluster of galaxies 
can be characterized by the fractal dimension $D\sim 2$\cite{pietronero}.
In cosmological simulations 
based on the standard cold dark matter scenario,
the density profile shows a power law distribution
(at outer region, $n\sim 3$ and 
at inner region, $n\sim 1.0-1.5$\cite{NFW97,makino01}.). 

Recently, 
in order to study the statistical properties of a self-gravitating system,
we proposed the self-gravitating ring model\cite{sota},
where all particles are moving, 
on a circular ring located in three-dimensional space,  
with mutual interaction of gravity in three-dimensional space.
The numerical simulation shows that
the system at the intermediate energy scale, 
where the specific heat becomes negative, 
has some peculiar properties such as 
non-Gaussian and power law velocity distribution($f(v)\sim v^{-2}$),
scaling mass distribution, and self-similar recurrent motion.
In this model, 
the {\it halo} particles which belong to the intermediate energy scale 
are considered to play an important role 
in realizing such specific characters.

We are interested in the origin of these scaling properties 
from statistical mechanical point of view. 
In order to study the statistical properties of long range interaction 
such as gravity, Ising model, and spin glass,
the model with power law potential has been used and 
revealed anomalous properties\cite{Ispolatov2001,Campa2001,Campa2002}. 
For example, 
a gravitational-like phase transition\cite{Ispolatov2001},  
reduction of mixing\cite{Campa2001}, 
and long relaxation\cite{Campa2002} are observed.
Using a model with an attractive $1/r^\alpha$ potential 
in general D-dimensional space,
we can control the extensivity of the system 
and the specific heat by changing the spatial dimension $D$ and 
the exponent of inverse power of the potential $\alpha$.

In this paper, 
we study the quasi-equilibrium state of N-body system 
with a power law potential.
As a first step, 
we consider the collisionless Boltzmann equation (CBE)
in place of N-body system and 
derive the self-similar stationary solution 
of CBE which has a scaling property appearing 
in the quasi-equilibrium state and discuss the linear stability 
by the use of energy functional analysis\cite{kandrup90,kandrup91,perez96,Rey96}.

In section \ref{sec:N}, 
we show some general properties of N-body systems 
with power law potential.
In section \ref{sec:SS}, 
we derive the self-similar stationary solution of CBE 
with an attractive $1/r^\alpha$ potential 
assuming spherical symmetry and isotropic orbit case 
in D-dimensional space.
Stability for the linear perturbation 
around the self-similar stationary solution is investigated 
in section \ref{sec:STABLE}.
Section \ref{sec:DISS} is devoted to discussion.

\section{\label{sec:N}N-body system with power law potential}

In this section,
we show some general characteristic properties of N-body system 
with power law potential.

We write the Hamiltonian for the N-body system 
with power law potential in the form:
\bea
 && H =
    \sum_{i=1}^N \frac{P_i^2}{2m}
    -\sum_{i<j}^N\frac{G m^2}{r_{ij}^{\alpha}},
\label{H}
\eea
where $r_{ij}:=|r_i-r_j|$ and 
$\alpha$ controls the range of interaction.

In this system, 
the virial equilibrium condition becomes
\bea
 2<K> + \alpha <\Phi> = 0,
\label{viri-con}
\eea
where $<K>$ is the time averaged kinetic energy and 
$<\Phi>$ is the time averaged potential energy.
{}From the expression $H=K+\Phi$, we have
\bea
 H = -\frac{2-\alpha}{\alpha}<K> = \frac{2-\alpha}{2}<\Phi>.
\label{viri-EKV}
\eea
{}From Eq.(\ref{viri-EKV}), 
the signature of the specific heat is determined 
by the signature of the term $-(2-\alpha)/\alpha$.

In order to clarify the extensivity / nonextensivity of the system, 
we focus on the $N$ dependence of the potential energy $\Phi$ 
under fixing the number density $N/L^D$\cite{Tsallis95}:
\bea
&&\frac{\Phi}{N} \sim 
\int^{N^{(1/D)}}drr^{D-1}r^{-\alpha} \sim N^{1-\alpha/D}.
\label{extens}
\eea
If the $N$ dependence of the potential energy per particle 
disappears for $N \rightarrow \infty$.
we define the system extensive.
Otherwise, we define the system nonextensive.
In the case of the gravity in $D$-dimensional space, 
since $\alpha=D-2$, the system is always nonextensive.
We summarize the signature of the specific heat of the system 
and the extensivity in Table.\ref{tab:dimension}.

\begin{table}
\caption{\label{tab:dimension}
In the system with an attractive $1/r^\alpha$ potential 
in $D$-dimensional space, 
the property of the specific heat and the extensivity is shown.
}
\begin{center}
\begin{tabular}{ccc}\hline
              & $0<\alpha<2$ & $\alpha>2$\\\hline
specific heat & negative     & positive  \\\hline
             & $\alpha<D$    & $\alpha>D$ \\\hline
extensivity & nonextensive & extensive  \\\hline
\end{tabular}
\end{center}
\end{table}%

\section{\label{sec:SS}Self-Similar stationary solution 
in Collisionless Boltzmann Equation (CBE)}

In this section, 
we derive a self-similar stationary solution 
in the collisionless Boltzmann equation (CBE);
\bea
 \frac{df}{dt}= \frac{\partial f}{\partial t}+[f,H]=0.
\eea
where $f=f(\bm{x},\bm{v},t)$ is a mass distribution function and 
$[A,B]$ denotes the Poisson bracket.

The stationary solution $f_0$ satisfies the following equation,
\bea
 [f_0,H]=\sum_{i=1}^{2D}\frac{\partial w^i}{\partial t}
                        \frac{\partial f_0}{\partial w^i}=0,
\label{CBE-S}
\eea
where $w^i=\{\bm{x},\bm{v}\}$.

For the coupled CBE and Poisson equation, 
R.N.~Henriksen and L.M.~Widrow\cite{SSS} studied 
the self-similar stationary solution 
in CBE with spherical symmetry in three-dimensional space
by applying the systematic method 
which is based on the work of B.~Carter and R.N.~Henriksen\cite{SS}.

Following R.N.~Henriksen and L.M.~Widrow\cite{SSS},
we study the self-similar stationary solution
with spherical symmetry and isotropic orbit case 
in general D-dimensional space.
The case that $D=3$ and $\alpha=1$ reduces to the work 
by R.N.~Henriksen and L.M.~Widrow\cite{SSS}.
By the generalization of the spatial dimension $D$ 
and the exponent of power of potential $\alpha$, 
we intend to investigate the relation between 
the extensivity of the system 
and the self-similarity.

{}From Eq.(\ref{CBE-S}), 
mass distribution function $f(r,v)$ obeys 
\bea
 v\partial_r f-\partial_r\Phi\partial_v f=0,
\label{Vlasov}
\eea
where $v:=\sqrt{v_r^2+v_\theta^2+v_\phi^2}$ and 
$\Phi$ is a potential.
The potential $\Phi$ satisfies the following equation,
\bea
 \frac{1}{r^{D-1}}\partial_r\left(r^{\alpha+1}\partial_r\Phi\right)
  = S_D^2 G\int v^{D-1}fdv, 
\label{Poisson}
\eea
where $S_D := 2\pi^{D/2}/\Gamma(D/2)$.
In the case of $\alpha=D-2$,
the above equation corresponds to Poisson equation.

A self-similar stationary solution satisfies the following equation,
\bea
 {\cal L}_{\bm k} f = 0,
\label{SS-con}
\eea
where
\bea
 {\cal L}_{\bm k} 
   := k^i\partial_i
    = \delta r\partial_r + \nu v\partial_v +\mu m\partial_m
\label{SS-def}
\eea
is a Lie derivative with respect to the vector $\bm k$ in phase space, 
and $\delta$, $\nu$, and $\mu$ are arbitrary constants.

In a dimensional space of length, velocity, and mass, 
we introduce vectors ${\bm a}=(\delta,\nu,\mu)$ and
${\bm d}_f$.
The vector ${\bm a}=(\delta,\nu,\mu)$
describes changes in the logarithms of dimensional quantities.
Each dimensional quantity $f$ in the problem has 
its dimension represented by the vector ${\bm d}_f$.
Using these vectors ${\bm a}$ and ${\bm d}_f$, 
the action of ${\bm k}$ reads
\bea
 {\cal L}_{\bm k}f = ({\bm d}_f\cdot {\bm a})f.
\label{SS-ad}
\eea

The dimensional quantities in current problem $f$, $\Phi$, and $G$
have the following dimensional covectors,
\bea
 {\bm d}_f    &=& (-D, -D, 1),\nonumber\\
 {\bm d}_\Phi &=& (0, 2, 0), \label{d-SD}\\
 {\bm d}_G    &=& (\alpha, 2, -1).\nonumber
\eea
The requirement of the invariance of $G$ 
under the rescaling group action (\ref{SS-def})
implies ${\bm d}_G\cdot {\bm a}=0$,
\bea
 \mu =\alpha\delta+2\nu.
\eea

The dimensional space is reduced to the subspace of
(length, velocity), wherein the rescaling group element
${\bm a}=(\delta, \nu)$ and 
\bea
 {\bm d}_f    &=& (\alpha-D, 2-D),\nonumber\\
 {\bm d}_\Phi &=& (0, 2).
\eea

Here we define the new coordinate $R(r)$ and $X$ 
in replacement of the original coordinate $r$ and $v$ such that 
\bea
 {\cal L}_{\bm k}R = 1,
\label{SS-R-con}\\
 {\cal L}_{\bm k}X = 0.
\label{SS-X-con}
\eea
From Eqs.(\ref{SS-R-con}) and (\ref{SS-X-con}), 
we choose 
\bea
 r|\delta| &=& e^{\delta R},
\label{SS-R}\\
 v &=& Xe^{\nu R}.
\label{SS-X} 
\eea
The transformation from the original coordinate $(r,v)$ to 
the self-similar coordinate $(R,X)$ is shown in Appendix\ref{sec:RX}.

Under the new coordinate, 
these physical quantities $f$ and $\Phi$ can be written in the form:
\bea
 f(X,R)    &=& \overline{f}(X)e^{-[(D-\alpha)\delta+(D-2)\nu]R},
\label{fXR}\\
 \Phi(X,R) &=& \overline{\Phi}(X)e^{2\nu R}.
\label{PhiXR}
\eea

Substituting Eqs.(\ref{fXR}) and (\ref{PhiXR}) 
into Eqs.(\ref{Vlasov}) and (\ref{Poisson}), 
these equations for a bounded solution yield
\bea
&&\!\!\!\!\!\!\!\!\!\!\!\! 
  \frac{d\ln \overline{f}}{d\ln X} 
  = -\frac{\left[D-2+(D-\alpha)\frac{\delta}{\nu}\right]X^2}
            {X^2+2\overline{\Phi}},
\label{Vlasov-2}\\
&& \!\!\!\!\!\!\!\!\!\!\!\!
 2\nu^2|\delta|^{D-\alpha-2}
 \left[2+\frac{\alpha\delta}{\nu}\right]\overline{\Phi}
  = S_D^2 G\int_0^{\sqrt{-2\overline{\Phi}}}
      \!\!\!\!\!\!\!\! X^{D-1}\overline{f}dX.
\label{Poisson-2}
\eea
Without loss of generality, we can set $\nu=1$.

Solving Eqs.(\ref{Vlasov-2}) and (\ref{Poisson-2}),
we have the following solution,
\bea
 \overline{f} = C|X^2+2\overline{\Phi}|^{-[(D-\alpha)\delta+(D-2)]/2},
\label{S-SS}
\eea
where
\bea
 C = \frac{\left|2+\alpha\delta\right||\delta|^{D-\alpha-2}
     \Gamma(D/2)\Gamma(2+(\alpha-D)\delta/2)}
     {2\pi^D G|-2\overline{\Phi}|^{(\alpha-D)\delta/2}
     \Gamma([4-D+(\alpha-D)\delta-D]/2)},\nonumber\\
\label{C-SSSD}
\eea
and  the following condition must be satisfied,
\bea
 (D-\alpha)\delta< 4-D.
\label{S-If}
\eea
Since $\overline{\Phi}<0$, 
from Eq.(\ref{Poisson-2}) we obtain the additional condition,
\bea
 \alpha\delta < -2.
\label{S-bound}
\eea

If these condition Eqs.(\ref{S-If}) and (\ref{S-bound}) 
are actually satisfied, 
we have the bounded self-similar stationary solution 
(\ref{S-SS}) and (\ref{C-SSSD}).
The mass distribution function $f$, the mass density $\rho$, 
and the velocity distribution $f(v)$ become respectively
\bea
 f(r,v) &=& C|2E|^{-[(D-\alpha)\delta+(D-2)]/2},
\label{f-SSD}\\
 \rho   &:=& S_D\int dv v^{D-1} f(r,v) \sim r^{\alpha-D+2/\delta},\\
 f(v)   &:= & S_D\int dr r^{D-1}f(r,v) \sim v^{\alpha\delta+2-D},
\eea
where $E$ denotes the mean field energy:
\bea
 E:=\frac{1}{2}v^2+\Phi_0.
\eea

The ratio of the average of the kinetic energy to the potential energy 
is as follows:
\bea
  \frac{<\Phi_0>}{<K>} &=& \frac{(D-\alpha)\delta-4}{2D}.
\label{DSSS-KP}
\eea
Since the solution (\ref{f-SSD}) we obtained is a bounded solution, 
which satisfies the following condition,
\bea
  \frac{(D-\alpha)\delta-4}{2D}&<&-1,
\label{bound}
\eea
the specific heat of the self-similar stationary solution 
is always negative.
The $\delta_*$ corresponds to a virial equilibrium state:
\bea
  \delta_* &=& \left\{
             \begin{array}{@{\,}ll}
                 -\frac{4}{\alpha}& (\alpha\neq D)\\
                 \mbox{arbitrary}& (\alpha=D).
             \end{array}                 
                \right.
\label{DSSS-VIRIAL}
\eea
If $(D-\alpha)(\delta -\delta_*)<0$,
the potential energy in this state is more dominant than 
in the virialized state.

The relation between pressure $P$ and mass density $\rho$ 
can be written in the form:
\bea
 P \sim \rho^{1+\frac{1}{1+(\alpha-D)\delta/2}}.
\eea
The above equation of state corresponds to Polytropes gas 
when identifying the Polytropes index $n$ as $1+(\alpha-D)\delta/2$.
Note that for $\alpha=D$, 
there is no self-similar stationary solution 
corresponding to the isothermal state.

As for gravity case ($\alpha=D-2$), 
the above solution (\ref{C-SSSD}) and (\ref{f-SSD}) in $D=3$ 
corresponds to the solution derived 
by R.N.~Henriksen and L.M.~Widrow\cite{SSS}.
For $D=1$ and $D=2$ where $\alpha=D-2\le 0$, 
we show the self-similar stationary solution 
in Appendix \ref{sec:D12}.

\section{\label{sec:STABLE}Linear perturbation analysis}

In this section, 
we investigate the stability of the self-similar stationary solution 
derived in the previous section for a symplectic linear perturbation, 
by energy functional analysis\cite{kandrup90,kandrup91,perez96,Rey96}.

As for the linear stability of the stationary solution 
in CBE of the gravity in three-dimensional space,
there has been much research
\cite{BT87,antonov61,antonov62,lyndenbell69,ipser74,
sygnet84,bartholomew71,kandrup90,kandrup91,perez96,Rey96}.
For the stationary state, assuming spherical symmetry, characterized 
by the mass distribution function $f_0$
specified as a function of the mean field energy $E$ and 
the squared angular momentum $J^2$,
if $\partial f_0/\partial E<0$ and $\partial f_0/\partial J^2<0$,
then the system is stable to the linear perturbation.

Following the work by J.~Perez and J.J.~Aly\cite{perez96} where
the stability of stationary solution 
in the coupled CBE and Poisson equation 
with spherical symmetry in three-dimensional space was studied,
we study the stability of the solution obtained in the previous section.

First, 
we explain a symplectic linear perturbation 
by energy functional analysis\cite{kandrup90,kandrup91,perez96,Rey96}.
In term of the mass distribution function $f(\bm{x}, \bm{v}, t)$, 
the Hamiltonian $H$ is written as follows,
\bea
&& H = \int d\Gamma \frac{\bm{v}^2}{2}f(\bm{x},\bm{v},t)
\nonumber \\
&& +\frac{G}{2}\int \!d\Gamma \!\int \!d\Gamma'\! 
     {\cal G}(|\bm{x}-\bm{x}'|)f(\bm{x},\bm{v},t)f(\bm{x}',\bm{v}',t),
\label{s-H}
\eea
where $d\Gamma:=d^D\bm{x}d^D\bm{v}$ is 
the $2D$-dimensional phase volume element
and the kernel ${\cal G}$ satisfies

\bea
 \Phi(\bm{x}) = 
   G\int d\Gamma'{\cal G}(|\bm{x}-\bm{x}'|)f(\bm{x}',\bm{v}',t).
\label{s-K}
\eea

We consider a small perturbation 
around the stationary solution $f_0$.
The distribution function and Hamiltonian can be expanded 
around the stationary solution as follows,
\bea
 f(\bm{x},\bm{v},t) &=& f_0 +\delta^{(1)}f+\delta^{(2)}f+\cdots ,
\label{s-pf}\\
 H        &=& H_0 +\delta^{(1)}H+\delta^{(2)}H+\cdots .
\label{s-pH}
\eea

Here we consider any symplectic perturbation, 
which can be generated from the stationary solution $f_0$ 
by use of a canonical transformation.
By using some generating function $K$, 
any symplectic deformation can be expressed in the form:
\bea
 f = e^{[K,\cdot]}f_0.
\label{s-sd}
\eea
{}From the above definition (\ref{s-sd}),
$f$ can also be expressed as follows,
\bea
 f = &&f_0 + [K,f_0]+\frac{1}{2!}[K,[K,f_0]]
\nonumber \\
   &&  +\frac{1}{3!}[K,[K,[K,f_0]]+\cdots.
\label{s-fe}
\eea

Introducing a small parameter $\epsilon$ 
which represents the amplitude of the perturbation,
we expand the generating function $K$ as
\bea
 K = \epsilon K^{(1)}+\epsilon^2 K^{(2)}+\epsilon^3 K^{(3)}+\cdots.
\label{s-Ke}
\eea
Further identifying $g^{(n)}=\epsilon^n K^{(n)}$, 
we obtain the perturbed quantities in the Eq.(\ref{s-pf}) 
in the form:
\bea
 \delta^{(1)}f &=& [g^{(1)},f_0],
\label{s-f1}\\
 \delta^{(2)}f &=& [g^{(2)},f_0]+\frac{1}{2}[g^{(1)},[g^{(1)},f_0]].
\label{s-f2}
\eea

The first order term in Eq.(\ref{s-pH}) becomes
\bea
 \delta^{(1)}H = \int d\Gamma E [g^{(1)},f_0],
\label{s-H1}
\eea
where $E$ is the energy of a particle,
\bea
 E := \frac{\bm{v}^2}{2}+\Phi_0,
\label{s-E}
\eea
where $\Phi_0$ is the potential energy generated by $f_0$.
Since $E$ and $f_0$ are conserved quantities, 
$\delta^{(1)}H=0.$

The next order term in Eq.(\ref{s-pH}) yields
\bea
&& \delta^{(2)}H 
   = \int d\Gamma E [g^{(2)},f_0]
       +\frac{1}{2}\int d\Gamma E[g^{(1)},[g^{(1)},f_0]]
\nonumber \\
 &&    +\frac{G}{2}\int d\Gamma \int d\Gamma'
       {\cal G}(|\bm{x}-\bm{x'}|)[g^{(1)},f_0][{g^{(1)}}',f_0'].
\label{s-H2}
\eea
The first term in Eq.(\ref{s-H2}) also vanishes 
and by the integration by parts, 
Eq.(\ref{s-H2}) is rewritten in the form:
\bea
&& \delta^{(2)}H 
   = -\frac{1}{2}\int d\Gamma [g^{(1)},f_0][g^{(1)},E]
\nonumber \\
&&   +\frac{G}{2}\int d\Gamma \int d\Gamma'
     {\cal G}(|\bm{x}-\bm{x'}|)[g^{(1)},f_0][{g^{(1)}}',f_0'].
\label{s-H2f}
\eea

Hereafter we consider the case that 
the stationary solution $f_0$ is a function of only the energy $E$.
In this case, we obtain
\bea
 [g^{(1)},f_0] &=& F_E[g^{(1)},E],
\label{s-gf}\\
 \int d^D\bm{v}[g^{(1)},f_0] 
   &=& \int d^D\bm{v}\partial_{\bm{x}}(F_E\bm{v}g^{(1)}),
\label{s-vgf}
\eea
where $F_E:=\partial_Ef_0$.

Integrating by parts and using Eqs.(\ref{s-gf}) and (\ref{s-vgf}), 
we have
\bea
&&\delta^{(2)}H = \nonumber \\
&&    \frac{1}{2}\int d\Gamma (-F_E)|[g^{(1)},E]|^2
      +\frac{G}{2}
     \int d\Gamma\partial_{\bm{x}}(-F_E \bm{v}g^{(1)})
\nonumber \\
&&    \times\int d\Gamma'\partial_{\bm{x'}} (-F'_E \bm{v}'{g^{(1)}}')
     {\cal G}(|\bm{x}-\bm{x}'|).
\label{s-H2n}
\eea

The linear perturbation $g^{(1)}$ has two kinds of gauge mode.
$(a)$ $g^{(1)}=g^{(1)}(E)$.
In this case, 
the linear perturbation of the mass distribution $\delta^{(1)}f$
is trivially zero.
$(b)$ $g^{(1)}=\bm{a}\bm{v}$ ($\bm{a}$ is a constant.).
This perturbation means the translation of the center of mass.
In order to consider the physical perturbation, 
we investigate the linear perturbation 
excluding the above gauge modes.

The stability for the linear perturbation\cite{bartholomew71,holm} 
reads:
\bea
\mbox{If $\delta^{(2)}H>0$, then the system is stable.}
\label{criterion}
\eea

\subsection{spherical mode}

Since the first order perturbed potential $\delta^{(1)}\Phi(r)$ 
satisfies
\bea
&& \frac{1}{r^{D-1}}
   \partial_r\left(r^{\alpha+1}\partial_r\delta^{(1)}\Phi(r)\right)
   = S_D G\int d^D\bm{v}' [{g^{(1)}}',f_0']\nonumber\\
   &&= S_D G\frac{1}{r^{D-1}}\partial_r\left(r^{D-1}
            \int d^D\bm{v}'F_E'v^{r'}{g^{(1)}}'\right),
\eea
the spatial derivative of $\delta^{(1)}\Phi(r)$ becomes
\bea
 \partial_r\delta^{(1)}\Phi(r)
   = \frac{S_DG}{r^{\alpha-D+2}}\int d^D\bm{v}' F_E'v^{r'}{g^{(1)}}'.
\label{1d-drphi}
\eea

{}From Eqs.(\ref{s-H2n}) and (\ref{1d-drphi}), 
we have
\bea
&& 2\delta^{(2)}H \nonumber\\
&& = \int d\Gamma (-F_E)|[g^{(1)},E]|^2
\nonumber\\
&&   -\int d\Gamma\partial_r(-F_E v^rg^{(1)})\delta^{(1)}\Phi(r)
\nonumber\\
&& = \int d\Gamma (-F_E)|[g^{(1)},E]|^2
     -S_D G\int \frac{d^D\bm{x}}{r^{\alpha-D+2}}
\nonumber \\
&&    \times
      \int d^D\bm{v}(-F_E vg^{(1)})\int d^D\bm{v}'(-F_E'v^{r'}{g^{(1)}}').
\label{1d-h2s}
\eea

Introducing new variables,
\bea
 g^{(1)}=:rv^r\mu(r,\bm{v},t),
\label{new-mu}
\eea
and using Schwartz's inequality, we have
\bea
 2\delta^{(2)}H 
   &=& \int d\Gamma (-F_E)|[\mu rv^r,E]|^2
       -GS_D\int \frac{d^D\bm{x}}{r^{\alpha-D+2}}
\nonumber \\
&&    \times\int d^D\bm{v}[-F_E r(v^r)^2\mu]
       \int d^D\bm{v'}[-F'_E r(v^{'r})^2\mu']\nonumber\\
   &\ge& \int d\Gamma (-F_E)|[\mu rv^r,E]|^2
         -GS_D\int \frac{d^D\bm{x}}{r^{\alpha-D+2}}
\nonumber \\
&&     \times\int d^D\bm{v}[-F_E r(v^r)^2\mu^2]
         \int d^D\bm{v'}[-F'_E r(v^{'r})^2]\nonumber\\
   &=& \int d\Gamma (-F_E)\left\{|[\mu rv^r,E]|^2
       -\frac{GS_D(rv^r)^2\mu^2\rho_0}{r^{\alpha-D+2}}\right\},
\nonumber\\&&
\label{SD-h2}
\eea
where $\rho_0$ is non-perturbed mass density:
\bea
 \rho_0 := \int d^D\bm{v}f_0
        = \int d^D\bm{v}(-F_E)(v^r)^2.
\eea

Using the property of the Poisson bracket, and 
the fact that the integral of the Poisson bracket over the phase space 
vanishes, the equation (\ref{SD-h2}) can be rewritten in the form:
\bea
 2\delta^{(2)}H 
   &\ge& \int d\Gamma(-F_E)\left\{|[\mu rv^r,E]|^2
       -\frac{GS_D(rv^r)^2\mu^2\rho_0}{r^{\alpha-D+2}}\right\}\nonumber\\
   &=& \int d\Gamma(-F_E)\left\{(rv^r)^2|[\mu,E]|^2+|\mu|^2|[rv^r,E]|^2
\right.\nonumber \\
&&\left.   +rv^r[\mu^2,E][rv^r,E]
           -\frac{GS_D(rv^r)^2\mu^2\rho_0}{r^{\alpha-D+2}}\right\}
\nonumber\\
   &=& \int d\Gamma(-F_E)\left\{(rv^r)^2|[\mu,E]|^2+|\mu|^2|[rv^r,E]|^2
\right.\nonumber \\
&&
        +[\mu^2rv^r[rv^r,E],E]-|\mu|^2rv^r[[rv^r,E],E]
\nonumber \\
&&\left.-|\mu|^2|[rv^r,E]|^2
        -\frac{GS_D(rv^r)^2\mu^2\rho_0}{r^{\alpha-D+2}}\right\}
\nonumber\\
   &=& \int d\Gamma(-F_E)\left\{(rv^r)^2|[\mu,E]|^2
\right.\nonumber \\
&&\left. -|\mu|^2rv^r[[rv^r,E],E]
        -\frac{GS_D(rv^r)^2\mu^2\rho_0}{r^{\alpha-D+2}}\right\}.
\eea

Using the following relation,
\bea
 [[rv^r,E],E] &=& -rv^r\left(\frac{d^2\Phi_0}{dr^2}
                  +\frac{3}{r}\frac{d\Phi_0}{dr}\right)\nonumber\\
              &=& -rv^r\left(\frac{GS_D\rho_0}{r^{\alpha-D+2}}
                  +\frac{2-\alpha}{r}
                  \frac{d\Phi_0}{dr}\right),
\eea
we obtain the final expression in the form:
\bea
&& \delta^{(2)}H \ge \nonumber \\
&&        \frac{1}{2}\int d\Gamma(-F_E)(rv^r)^2\left(|[\mu,E]|^2
           +|\mu|^2\frac{2-\alpha}{r}\frac{d\Phi_0}{dr}\right).
\nonumber\\
\label{S-smode}
\eea
{}From Eqs.(\ref{s-f1}), (\ref{s-H2f}), and (\ref{S-smode}), 
if $\delta^{(2)}H=0$ when $F_E<0$ and $\alpha\le 2$, 
$\delta^{(1)}f=0$.
Since this is a gauge mode, 
we conclude that
\bea
\mbox{If $F_E<0$ and $\alpha\le 2$, then $\delta^{(2)}H>0$.}
\label{S-criterion}
\eea

{}From the self-similar stationary solution Eq.(\ref{S-SS}), 
we have
\bea
 F_E = \mbox{sgn}(E)
       [(\alpha-D)\delta+2-D]C|2E|^{[(\alpha-D)\delta-D]/2}.
\label{S-fe}
\eea

\begin{figure}
\includegraphics[width=8cm]{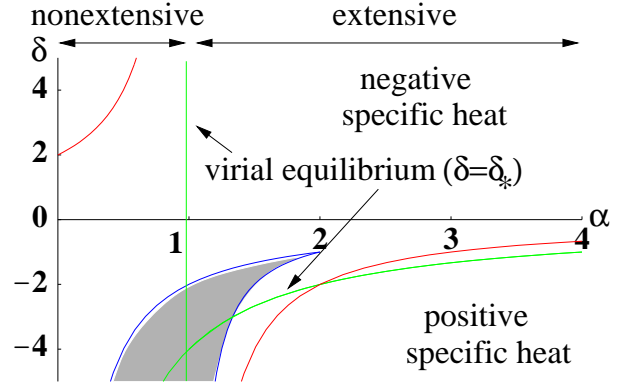}
\caption{\label{fig:a1}
Stability chart for the one-dimensional case ($D=1$).
The dark region corresponds to
the stable region (\ref{a1-stable}) 
in the parameter space ($\delta$, $\alpha$).
The property of the specific heat 
and the extensivity also are shown.}
\end{figure}

%
%
As an explicit example, 
we consider $D=1$ case.
{}From Eqs.(\ref{S-If}), (\ref{S-bound}), (\ref{bound}), and (\ref{S-criterion}), 
if the following condition is satisfied, 
the self-similar stationary solution Eq.(\ref{f-SSD}) is stable.

\bea
-\frac{1}{\alpha-1}<\delta<-\frac{2}{\alpha}
&\quad\quad\mbox{($1<\alpha\le 2$)} \nonumber\\
\delta < -\frac{2}{\alpha} 
&\quad\quad\mbox{($0<\alpha\le1$)}
\label{a1-stable}
\eea
In Fig.\ref{fig:a1}, 
we show the region where exits the stable self-similar stationary solution 
in the parameter space $(\alpha, \delta)$.

Note that 
in the above calculation, 
we use the integration by parts and neglect the surface term at the origin.
Since the self-similar stationary solution obtained in this paper is 
singular at the boundary, 
the surface term can not be neglected in general.
In the realistic situation, however, 
the self-similarity appears in the intermediate scale
since the system has a cut off scale in the short distance. 
We suppose that the self-similar stationary solution 
can be connected with some regular solution near the boundary 
by regularization such as $r \rightarrow (r^2+a^2)^{1/2}$ 
where $a$ is a cut off scale,
and the boundary term can be neglected.

\subsection{aspherical mode}

Next, we consider the aspherical mode.
Since it is difficult to analyze a general case, 
we study the gravity case in D-dimensional space ($\alpha=D-2$).

By the integral of Poisson equation over the configuration space 
and integration by parts, 
we have
\bea
 -\frac{1}{S_D}\int d^D\bm{x}|\nabla\delta^{(1)}\Phi|^2
  = \int d\Gamma \delta^{(1)}f\delta^{(1)}\Phi,
\label{dpdp-fp}
\eea
where 
\bea
 \delta^{(1)}\Phi:=\int d^D\bm{v}{\cal G}\delta^{(1)}f.
\label{def-phi1}
\eea

{}From Eqs.(\ref{s-H2f}), (\ref{dpdp-fp}), and (\ref{def-phi1}),
we have
\bea
 \delta^{(2)}H 
   = \frac{1}{2}\int d\Gamma \frac{(\delta^{(1)}f)^2}{-F_E}
     -\frac{G}{2S_D}\int d^D\bm{x}|\nabla\delta^{(1)}\Phi|^2.
\label{as-H2}
\eea

Here we introduce $\delta^{(1)}\tilde{f}$ as follows,
\bea
 \delta^{(1)}f =: F_E \delta^{(1)}\Phi+\delta^{(1)}\tilde{f}.
\label{def-tf1}
\eea

Substituting Eq.(\ref{def-tf1}) into Eq.(\ref{as-H2}), 
we obtain
\bea
&& \delta^{(2)}H 
\nonumber \\
&&   = \frac{1}{2}\int d\Gamma \left[
     \frac{(\delta^{(1)}\tilde{f})^2}{-F_E}
     +F_E(\delta^{(1)}\Phi)^2-2\delta^{(1)}f\delta^{(1)}\Phi\right]
\nonumber \\
&&  -\frac{G}{2S_D}\int d^D\bm{x}|\nabla\delta^{(1)}\Phi|^2,
\nonumber\\
&&   = \frac{1}{2}\int d\Gamma \frac{(\delta^{(1)}\tilde{f})^2}{-F_E}
       +\frac{1}{2S_D}\int d^D\bm{x}\biggl\{|\nabla\delta^{(1)}\Phi|^2
\nonumber \\
&&   -S_D\left[
     \int d^D\bm{v}(-F_E)|\delta^{(1)}\Phi|^2\right]\biggr\}.
\label{as-H2p}
\eea

Moreover, using the new variable $w$ which is defined by 
\bea
 \delta^{(1)}\Phi =: w(\bm{x},t)\partial_r\Phi_0,
\eea
we can rewrite the equation (\ref{as-H2p}) in the form:
\bea
&& \delta^{(2)}H \nonumber \\
&& = \frac{1}{2}\int d\Gamma \frac{(\delta^{(1)}\tilde{f})^2}{-F_E}
\nonumber\\
&&   +\frac{1}{2S_D}\int d^D\bm{x}\Biggl\{(\partial_r\Phi_0)^2\left[
     |\nabla w|^2
     -S_D\int d^D\bm{v}(-F_E)|w|^2\right]
\nonumber \\
&&   -|w|^2\partial_r\Phi_0\nabla^2\partial_r\Phi_0\Biggr\}.
\label{as-H2w}
\eea

By straightforward calculation, 
we obtain
\bea
 \nabla^2\partial_r\Phi_0=
   S_D\int d^D\bm{v}F_E\partial_r\Phi_0+(D-1)\frac{\partial_r\Phi_0}{r^2}.
\label{as-rphi0}
\eea

By using Wirtinger's inequality, 
we have
\bea
 \int\left[|\nabla_s w|^2-\frac{D-1}{r^2}|w|^2\right]d\Omega
   \ge 0,
\label{Wir}
\eea
where $\nabla_s:=\nabla-\frac{\bm{r}}{|\bm{r}|}\frac{\partial}{\partial r}$.

{}From Eqs.(\ref{as-H2w}), (\ref{as-rphi0}), and (\ref{Wir}), 
we get the final expression in the form:
\bea
&& \delta^{(2)}H \nonumber\\
&&   = \frac{1}{2}\int d\Gamma \frac{(\delta^{(1)}\tilde{f})^2}{-F_E}
\nonumber\\
&&   +\frac{1}{2S_D}\int d^D\bm{x}(\partial_r\Phi_0)^2\left[
     |\partial_r w|^2+|\nabla_s w|^2-\frac{D-1}{r^2}|w|^2\right]
\nonumber \\
&&  \ge \frac{1}{2}\int d\Gamma \frac{(\delta^{(1)}\tilde{f})^2}{-F_E}
     +\frac{1}{2S_D}\int d^D\bm{x}(\partial_r\Phi_0)^2
     |\partial_r w|^2.
\label{as-H2f}
\eea

\begin{figure}
\includegraphics[width=8cm]{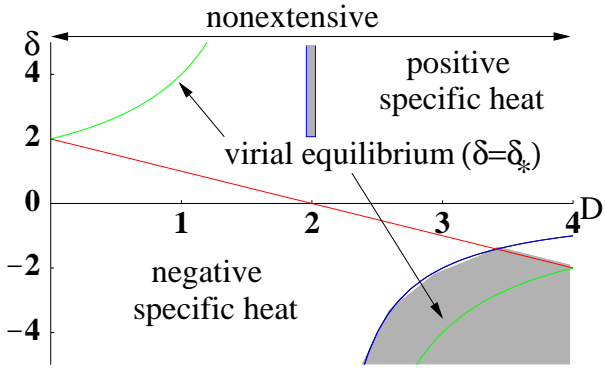}
\caption{\label{fig:SD}
Stability chart for the gravity case ($\alpha=D-2$).
The dark region corresponds to
the stable region (\ref{SSS-stable}) 
in the parameter space ($\delta$, $D$).
The property of the specific heat and 
the extensivity also are shown.
}
\end{figure}

{}From Eq.(\ref{as-H2f}), 
if $\delta^{(2)}H=0$ when $F_E<0$, 
then $\delta^{(1)}f=0$ or $g^{(1)}=\bm{a}\bm{v}$.
Since this is a gauge mode, 
we conclude that
\bea
\mbox{If $F_E<0$, then $\delta^{(2)}H>0$.}
\label{AS-criterion}
\eea
This condition is weaker than the condition (\ref{S-criterion}).
In the gravity case ($\alpha=D-2$), 
{}from Eqs.(\ref{S-If}), (\ref{S-bound}), (\ref{bound}), and (\ref{S-criterion}), 
if the following condition is satisfied, 
the self-similar stationary solution Eq.(\ref{f-SSD}) is stable.
\bea
  \delta_* &=& \left\{
             \begin{array}{@{\,}ll}
                 \delta < -\frac{2}{D-2}& (2<D\le 2+\sqrt{2})\\
                 \delta < 2-D           &  (2+\sqrt{2}<D \le 4).
             \end{array}                 
                \right.
\label{SSS-stable}
\eea
In Fig.\ref{fig:SD}, 
we show the region where exists the stable self-similar stationary solution 
in the parameter space $(D, \delta)$.
This stability condition (\ref{S-criterion}) is consistent with 
the work by J.~Perez and J.J.~Aly\cite{perez96}($\alpha=1,D=3$).

\section{\label{sec:DISS}Discussion}

We studied the self-similar stationary solution 
in collisionless Boltzmann equation 
with the attractive $1/r^\alpha$ potential. 
Assuming the spherical symmetric and isotropic orbit
in D-dimensional space, 
we investigate the linear stability of the solution.
In the above model, 
we can control the extensivity of the system 
and the signature of the specific heat
by changing the spatial dimension $D$ and 
the exponent of inverse power of the potential $\alpha$.

The self-similar stationary solution can be expressed 
in the form of the power law of the energy.
The exponent of the power is determined 
by the power of the potential $\alpha$, spatial dimension $D$, 
and the scaling parameter $\delta$.
Here we interpret $\delta$ as a parameter 
which denotes the virial ratio of the initial state.

By use of the energy functional approach,
we investigated the stability of the self-similar stationary solution
in terms of a symplectic linear perturbation.
As for the spherical symmetric and isotropic orbit 
of the gravity in D-dimensional space ($\alpha=D-2$), 
we found that the system is stable 
if the mass distribution function decreases monotonically and 
the spatial dimension is less than 4.
As for the power-law potential in one-dimensional space ($D=1$), 
we found that the system is stable 
if the mass distribution function decreases monotonically and 
the inverse power index of the potential is less than 2.
The self-gravitating ring model\cite{sota} 
is similar to the case of $\alpha=1$ in one-dimensional space.
{}From the form of the velocity distribution 
obtained by a numerical simulation, $\delta\sim -3$.
This case belongs to the stable self-similar stationary solution.

The stable self-similar stationary solution we obtained 
includes the virial equilibrium state 
in the case of $\delta=\delta_*$ 
As for the extensivity of the system, 
the nonextensive system has far more stable scaling solutions 
than the extensive system in the parameter space 
($\delta$, $\alpha$, $D$).

In the time evolution of the collisionless system, 
assuming the spherical symmetry and isothermal case, 
Larson-Penston solution which shows self-similar collapse 
is the attractor\cite{hanawa97}.
By such a self-similar time evolution of system, 
we expect that the class of 
the stable self-similar stationary solution obtained in this paper 
plays an important role as a quasi-equilibrium state 
with a long range interaction such as gravity.
In the realistic situation, 
since the anisotropic velocity space is important, 
we would like to extend this analysis to the anisotropic case 
in our future work.

\begin{acknowledgments}
We would like to thank Professor M.~Morikawa, Y.~Sota, and T.~Tatekawa
for useful discussions and comments.
\end{acknowledgments}

\appendix

\section{Transformation to self-similar coordinate}
\label{sec:RX}

Using the original coordinate $(r,v)$, 
the self-similar coordinate $(R,X)$ is defined by
\bea
 R &:=& \delta^{-1}\ln(r|\delta|),\\
 X &:=& ve^{-\nu R}.
\eea

The derivative with respect to the original coordinate can be
expressed by the self-similar coordinate in the form:
\bea
 \partial_r 
  &=& \left(\frac{\partial R}{\partial r}\right)\Big|_X\partial_R+
    \left(\frac{\partial X}{\partial r}\right)\Big|_R\partial_X,
\nonumber \\
  &=& \mbox{sgn}(\delta)e^{-\delta R}\left(\partial_R-\nu X\partial_X\right)\\
 \partial_v 
  &=& \left(\frac{\partial X}{\partial v}\right)\Big|_R\partial_X
  = e^{-\nu R}\partial_X.
\eea

\section{Stability condition for gravity case in $D=1$ and $2$ ($\alpha=D-2$)}
\label{sec:D12}

For the case that the potential $\overline{\Phi}$ is positive, 
Eq.(\ref{Poisson-2}) is modified as 
\bea
 2\nu^2\left[2+(D-2)\frac{\delta}{\nu}\right]\overline{\Phi}
  &=& S_D^2 G\int_0^{\infty}
      \!\!\!\!\!\!\!\! X^{D-1}\overline{f}dX.
\label{Poisson-2-D12}
\eea

\subsection{$D=1$ case}
\label{sec:D1}

Since the potential $\overline{\Phi}$ is positive, 
from Eq.(\ref{Poisson-2-D12}), 
\bea
 \delta &<& 2.
\label{S-bound-D1}
\eea

By integrating Eq.(\ref{Poisson-2-D12}),
we have a self-similar solution (\ref{S-SS}) and 
\bea
 C = \frac{\left|2-\delta\right|\Gamma(\delta-1/2)|2\overline{\Phi}|^\delta}
    {\sqrt{\pi} G\Gamma(\delta-1)},
\label{C-SSSD1}
\eea
if the following condition is satisfied:
\bea
 \delta &>& 1.
\label{S-IF-D1}
\eea

{}From Eqs. (\ref{S-bound-D1}), (\ref{S-IF-D1}), 
(\ref{S-criterion}) and (\ref{AS-criterion}), 
the stability condition for linear perturbation yields 
\bea
&& 1< \delta < 2.
\label{SSS-stable-D1}
\eea

The ratio of the average of the kinetic energy to the potential energy 
is the same as Eq.(\ref{DSSS-KP}) in $D=1$.
However, if $\delta\le 2$, 
the integral of the kinetic energy over the velocity space diverges.
For this reason, 
there do not exist a stable self-similar stationary solution 
in $D=1$.

\subsection{$D=2$ case}
\label{sec:D2}

If the potential $\overline{\Phi}$ is negative,
the condition that the bounded self-similar solution exists 
is the same as Eqs.(\ref{S-If}) and (\ref{S-bound}).
Since this case does not satisfy the condition (\ref{S-bound}),
the only case is that $\overline{\Phi}>0$.

In this case, 
from Eq.(\ref{Poisson-2-D12}), 
the condition that a bounded self-similar solution exists 
yields
\bea
 \delta &>& \frac{1}{2},
\label{S-IF-D2}
\eea
and the integral constant of (\ref{S-SS}) is
\bea
 C = \frac{2\Gamma(\delta)|2\overline{\Phi}|^\delta}
    {\pi^{5/2} G\Gamma(\delta-1/2)}.
\label{C-SSSD2}
\eea

{}From Eqs. (\ref{S-criterion}), (\ref{AS-criterion}), and (\ref{S-IF-D2}), 
the stability condition against linear perturbation yields 
\bea
\delta &>& \frac{1}{2}.
\label{SSS-stable-D2}
\eea

The ratio of the average of the kinetic energy to the potential energy 
is the same as Eq.(\ref{DSSS-KP}) in $D=2$.
However, similar to the $D=1$ case, if $\delta\le 2$, 
the integral of the kinetic energy over the velocity space diverges.
Finally, 
if $\delta > 2$, 
the self-similar stationary solution in $D=2$ is stable.
In this case, 
the specific heat is always positive.




\begin{thebibliography}{99}

\bibitem{larson81}
R.B.~Larson, 
Mon.~R.~astron.~Soc.{\bf 194} (1981), 809.

\bibitem{falgarone91}
E.~Falgarone, T.G.~Phillips, and C.K.~Walker, 
ApJ.{\bf 378} (1991), 186.

\bibitem{merritt96}
D.~Merritt and T.~Fridman, 
ApJ.{\bf 460} (1996), 136.

\bibitem{pietronero} 
F.S.~Labini, M.~Montuori, and L.~Pietronero, 
Phys.~Rep.{\bf 61} (1998), 293.

\bibitem{NFW97}
J.F.~Navarro, C.S.~Frenk, and S.D.M.~White, 
ApJ.{\bf 490} (1997), 493.

\bibitem{makino01}
T.~Fukushige and J.~Makino, 
ApJ.{\bf 557} (2001), 533.

\bibitem{sota} 
Y.~Sota, O.~Iguchi, M.~Morikawa, T.~Tatekawa, and K.~Maeda,
Phys.~Rev.~E {\bf 64} (2001), 056133.


\bibitem{Ispolatov2001} 
I.~Ispolatov and E.D.G.~Cohen, 
Phys.~Rev.~Lett. {\bf 87} (2001), 210601.

\bibitem{Campa2001} 
A.~Campa, A.~Giansanti, D.~Moroni, and C.~Tsallis,
Phys.~Lett.~A {\bf 286} (2001), 251.

\bibitem{Campa2002} 
A.~Campa, A.~Giansanti, and D.~Moroni,
Physica ~A {\bf 305} (2002), 137.

\bibitem{kandrup90}
H.E.~Kandrup,
ApJ.{\bf 351} (1990), 104.

\bibitem{kandrup91}
H.E.~Kandrup,
ApJ.{\bf 370} (1991), 312.

\bibitem{perez96}
J.~Perez and J.J.~Aly,
Mon.~Not.~R.~Astron.~Soc.{\bf 280} (1996), 689.

\bibitem{Rey96}
J.~Perez and M.~Lachieze-Rey,
ApJ.{\bf 465} (1996), 54.


\bibitem{Tsallis95} 
P.~Jund, S.G.~Kim, and C.~Tsallis,
Phys.~Rev.~B {\bf 52} (1995), 50.

\bibitem{SSS}
R.H.~Henriksen and L.M.Widrow, 
Mon.~Not.~R.~Astron.~Soc.{\bf 276} (1995), 679.

\bibitem{SS}
B.~Carter and R.H.~Henriksen, 
J.~Math.~Phys.{\bf 32} (1991), 2580.

\bibitem{BT87}
J.~Binney and S.~Tremaine,
{\em Galactic Dynamics.}\\
( Princeton Univ. Press, Princeton ) (1993).

\bibitem{antonov61}
V.A.~Antonov, 
Sov.~Astron.{\bf 4} (1961), 859.

\bibitem{antonov62}
V.A.~Antonov, 
Vestnik Leningrad Univ.{\bf 7} (1962), 135.

\bibitem{lyndenbell69}
D.~Lynden-bell and N.~Sannit,
Mon.~Not.~R.~Astron.~Soc.{\bf 143} (1969), 167.

\bibitem{ipser74}
J.R.~Ipser,
ApJ.{\bf 232} (1974), 863.

\bibitem{sygnet84}
J.F.~Sygnet, G.~Des~Forets, M.~Lachieze-Rey, and R.~Pellat,
ApJ.{\bf 276} (1984), 737.

\bibitem{bartholomew71}
P.~Bartholomew,
Mon.~Not.~R.~Astron.~Soc.{\bf 151} (1971), 333.

\bibitem{holm}
D.D.~Holm, J.E.~Marden, T.~Ratiu, and A.~Weinstein,
Phys.~Rep.{\bf 123}(1985), 1.

\bibitem{hanawa97}
T.~Hanawa and K.~Nakamura,
ApJ.{\bf 484} (1997), 238.

\end{thebibliography}
\end{document}